\documentclass[journal,twoside]{IEEEtran}
\usepackage{cite}
\usepackage{booktabs}
\usepackage{amsmath}
\usepackage{amssymb}
\usepackage{algorithm}
\usepackage{algpseudocode}

\usepackage{graphicx}
\graphicspath{{/Users/geyifan/Desktop/通感控/论文图片}}
\ifCLASSINFOpdf
\else
\fi
\usepackage{amsmath,amssymb,amsfonts}
\usepackage{bm}

\usepackage{array}

\ifCLASSOPTIONcompsoc
  \usepackage[caption=false,font=normalsize,labelfont=sf,textfont=sf]{subfig}
\else
  \usepackage[caption=false,font=footnotesize]{subfig}
\fi
\usepackage{fixltx2e}

\usepackage{stfloats}
\usepackage{url}

\title{Resilience in Wireless: A Lyapunov-Constrained Integrated Sensing, Communication, and Control Approach}
\author{Yifan~Ge,
        Ningyan~Guo,~\IEEEmembership{Member,~IEEE},
        Yuanhao~Cui,~\IEEEmembership{Member,~IEEE},
        Dingyou~Ma,~\IEEEmembership{Member,~IEEE},
        Zhiyong~Feng,~\IEEEmembership{Senior~Member,~IEEE},
        and Yonina~C.~Eldar,~\IEEEmembership{Fellow,~IEEE}%
        \thanks{This work was supported by the Fundamental and Interdisciplinary Disciplines Breakthrough Plan of the Ministry of Education of China under Grant JYB2025XDXM119.}

\thanks{Yifan Ge, Ningyan Guo, Yuanhao Cui, Dingyou Ma and Zhiyong Feng are with the School of Information and Communication Engineering, Beijing University of Posts and Telecommunications, Beijing 100876, China.}
\thanks{Yonina C. Eldar is with the Faculty of Mathematics and
Computer Science, Weizmann Institute of Science, Rehovot 7610001,
Israel.}
\thanks{Corresponding authors: Ningyan Guo (guoningyan@bupt.edu.cn) and Yuanhao Cui (yuanhao.cui@bupt.edu.cn).}
}

\begin{document}

\maketitle

\begin{abstract} This paper investigates resilience in wireless through state-aware physical-layer resource allocation in an integrated sensing, communication, and control (ISCC) system comprising a single base station (BS) and a single controlled agent. Conventional beamforming designs typically employ fixed link-quality targets and therefore do not account for how communication and sensing requirements change with the physical state. To address this limitation, we develop a Lyapunov-constrained ISCC framework that incorporates instantaneous control risk into the shared beamforming design. The resulting state-dependent constraint becomes tighter as the system error increases and relaxes as the system approaches the desired state, thereby adapting the transmit power and beamforming vector to the physical-task condition. The BS estimates the physical state from radar echoes, and the resulting beam-dependent sensing uncertainty is incorporated into the Lyapunov constraint. This formulation couples packet-delivery reliability and sensing accuracy within the same physical-layer beamforming decision. Under ideal sensing, the resulting problem admits an equivalent convex formulation and a closed-form solution. Under practical sensing uncertainty, the problem becomes non-convex, for which we develop an algorithm based on semidefinite relaxation (SDR) and successive convex approximation (SCA), together with rank-one recovery. We establish the mean-square ultimate boundedness of the closed-loop state under the stated assumptions and prove that every solvable SCA subproblem admits a rank-one optimal solution. Numerical results show a small empirical rank-one recovery loss. Simulations demonstrate that the proposed state-aware beamforming design supports closed-loop recovery from a high-risk state toward a bounded operating region, with a lower peak linear-quadratic regulator (LQR) cost and faster system-error decay than the static ISAC baseline. Relative to the control-aware baseline, it achieves a lower LQR cost as the sensing uncertainty increases, at the expense of higher transmit power. \end{abstract}


%
\IEEEpeerreviewmaketitle

\section{Introduction}
\subsection{Motivation}\IEEEPARstart{C}{onventional} physical-layer beamforming designs typically rely on fixed link-quality targets, such as the bit error rate, outage probability, and communication success probability~\cite{Kabadurmus2018Reliability}. These metrics characterize transmission reliability under given link conditions, but they do not capture how the communication and sensing requirements of a physical task change under disturbances. Addressing this limitation is consistent with the IMT-2030 vision, which identifies security and resilience as overarching design principles and defines resilience as the capability of networks and systems to continue operating correctly during and after disruptions~\cite{ITU2023IMT2030}. In line with this vision, recent studies have emphasized that wireless systems should adapt their operation to disturbances and unexpected changes rather than rely solely on static reliability targets~\cite{Sterbenz2010Resilience,Smith2011NetworkResilience,Bennis_Resilient_2025,Khaloopour_Resilience_2024}. Within this broad vision, this paper considers resilience in wireless from a state-aware physical-layer resource-allocation perspective rather than as a general network-level property. Specifically, we study a single BS serving one controlled agent, where a shared beam is used for packet delivery and physical-state sensing. The BS combines channel state information (CSI) with radar echoes to adapt the packet-delivery reliability requirement and beamforming vector to the instantaneous physical state. However, jointly incorporating physical-state evolution and beam-dependent sensing uncertainty into wireless beamforming remains insufficiently studied. To address this gap, we develop a Lyapunov-constrained ISCC framework for state-aware physical-layer beamforming.

\subsection{Related Work}

Early reliability-oriented beamforming imposed fixed signal-to-interference-plus-noise ratio (SINR) constraints. Under perfect CSI, Wiesel \textit{et al.} minimized the downlink transmit power subject to user-specific SINR requirements through conic optimization~\cite{Wiesel2006Precoding}. Ultra-reliable and low-latency communications (URLLC) subsequently shifted the design objective toward packet error probability and latency, for which short-packet effects cannot be accurately characterized by Shannon capacity alone. Popovski \textit{et al.} examined packet structure, access protocols, and diversity for reliable low-latency transmission~\cite{Popovski2018URLLC}, while Bennis \textit{et al.} considered rare events, latency tails, and risk in large wireless systems~\cite{Bennis_URLLC_2018}. Using the finite-blocklength model, He \textit{et al.} jointly optimized multiuser downlink beamforming and transmit power under short-packet rate and reliability requirements~\cite{He2021URLLCBeamforming}. Robust beamforming further addressed CSI uncertainty: Shenouda and Davidson considered worst-case QoS constraints under bounded CSI errors~\cite{Shenouda2007Robust}, Song \textit{et al.} analyzed rank-one recovery for SDR-based designs~\cite{Song2012RobustBeamforming}, and Wang \textit{et al.} developed convex approximations for chance-constrained outage requirements~\cite{Wang2014Outage}. Despite this progression from nominal SINR constraints to short-packet and robust reliability, the required reliability remains predefined and independent of the physical state.

At the network layer, Lyapunov methods have been widely used to guide wireless resource allocation. Tassiulas and Ephremides derived max-weight scheduling by minimizing an upper bound on quadratic queue drift, thereby stabilizing arrival rates within the network capacity region~\cite{Tassiulas1992MaxWeight}. Neely \textit{et al.} extended this approach to time-varying wireless networks through joint routing and power allocation based on queue and channel states~\cite{Neely2005Dynamic}, while Georgiadis \textit{et al.} incorporated communication utility and cost through the drift-plus-penalty framework, yielding per-slot resource-allocation decisions~\cite{Georgiadis2006CrossLayer}. Subsequent control-aware studies connected wireless allocation with physical-state evolution. Gatsis \textit{et al.} jointly considered the LQR cost and transmit power through a Markov decision process~\cite{Gatsis_TAC_2014}, and later minimized the expected transmit power subject to an expected decrease condition on a quadratic Lyapunov function~\cite{Gatsis2015Opportunistic}. Eisen \textit{et al.} derived a state-dependent packet-delivery requirement from a one-step Lyapunov condition and optimized frequency allocation, transmission rate, and scheduling~\cite{Eisen2019IoTJ}, whereas Deng \textit{et al.} expressed mean-square stability as linear inequalities for joint control and network scheduling~\cite{Deng_WNCS_2021}. These methods adapt wireless resources to physical states, but mainly optimize power, channel access, or scheduling under available state information, without accounting for beam-dependent sensing uncertainty. Joint packet delivery and state sensing through a shared beam therefore remain insufficiently studied.

Integrated sensing and communication (ISAC) enables shared beamforming for wireless transmission and target sensing~\cite{Liu_TSP_2022,Liu_TWC_2018}. Recent surveys review ISAC foundations, physical-layer designs, networking aspects, and open challenges~\cite{Lu2024ISACChallenges}, and envision multifunctional perceptive networks integrating wireless communication with environmental sensing~\cite{Cui2025PerceptiveNetwork}. At the physical layer, existing ISAC beamforming designs typically optimize communication and sensing under predefined SINR and Cram\'er--Rao bound (CRB) requirements. Beyond conventional ISAC, recent work has investigated the co-design of sensing, communications, and control for low-altitude wireless networks, where localization and LQR performance are jointly considered through resource allocation, UAV deployment, and user scheduling~\cite{Jin2025LowAltitudeSC2}. However, these designs do not explicitly couple packet delivery and beam-dependent sensing uncertainty through a state-dependent Lyapunov constraint.

In summary, existing methods either optimize predefined link-level requirements or use physical-state information to adapt communication resources. Although recent ISCC studies incorporate control objectives, the joint effect of packet delivery and beam-dependent sensing uncertainty has not been explicitly translated into a state-dependent physical-layer beamforming constraint. This gap motivates the proposed framework.
\subsection{Contributions}

Overall, we propose a Lyapunov-constrained ISCC beamforming framework for state-aware physical-layer resource allocation. The main contributions are summarized as follows.

\begin{itemize}

\item We develop a Lyapunov-constrained ISCC framework for state-aware
physical-layer beamforming. The BS uses a shared beam for packet delivery and physical-state sensing. Unlike designs with fixed QoS targets, the proposed method adapts the transmit power and beamforming vector to the instantaneous physical state. 

\item We incorporate a state-dependent Lyapunov constraint into the physical-layer beamforming optimization. The constraint captures the joint effects of packet losses, sensing errors, and process noise on physical-state evolution. We further prove the mean-square ultimate boundedness of the closed-loop state under the stated system and sensing assumptions.

\item We first study the ideal-sensing case, where the beamforming problem admits an equivalent convex formulation and a closed-form solution. We then extend the proposed framework to practical sensing with state estimation errors. In this case, the coupling between communication success probability and sensing accuracy leads to a non-convex problem. We develop an SDR-SCA algorithm that converges to a stationary solution under standard conditions. We further prove that every solvable semidefinite programming (SDP) subproblem admits a rank-one optimal solution, showing that SDR is tight at each SCA iteration.

\end{itemize}


\subsection{Organization and Notations}

The remainder of this paper is organized as follows. Section II presents the communication and control models. Section III formulates the state-dependent beamforming problem under ideal and practical sensing. Section IV develops the SDR-SCA solution, and Section V presents the simulation results. Section VI concludes the paper.\textit{Notations:} Scalars are denoted by italic letters, vectors by boldface lowercase letters, and matrices by boldface uppercase letters. The superscripts $(\cdot)^T$ and $(\cdot)^H$ denote the transpose and conjugate transpose, respectively. $\operatorname{Tr}(\cdot)$, $\operatorname{rank}(\cdot)$, and $\mathbb{E}[\cdot]$ denote the trace, rank, and expectation operators, respectively. The notation $\|\cdot\|$ denotes the Euclidean norm for vectors and the Frobenius norm for matrices. $\mathbb{C}^{M\times N}$ denotes the space of $M\times N$ complex-valued matrices, and $\mathbf{X}\succeq\mathbf{0}$ means that $\mathbf{X}$ is positive semidefinite.

\section{System Model}
\begin{figure*}[!t]
    \centering
    \includegraphics[width=0.95\textwidth]{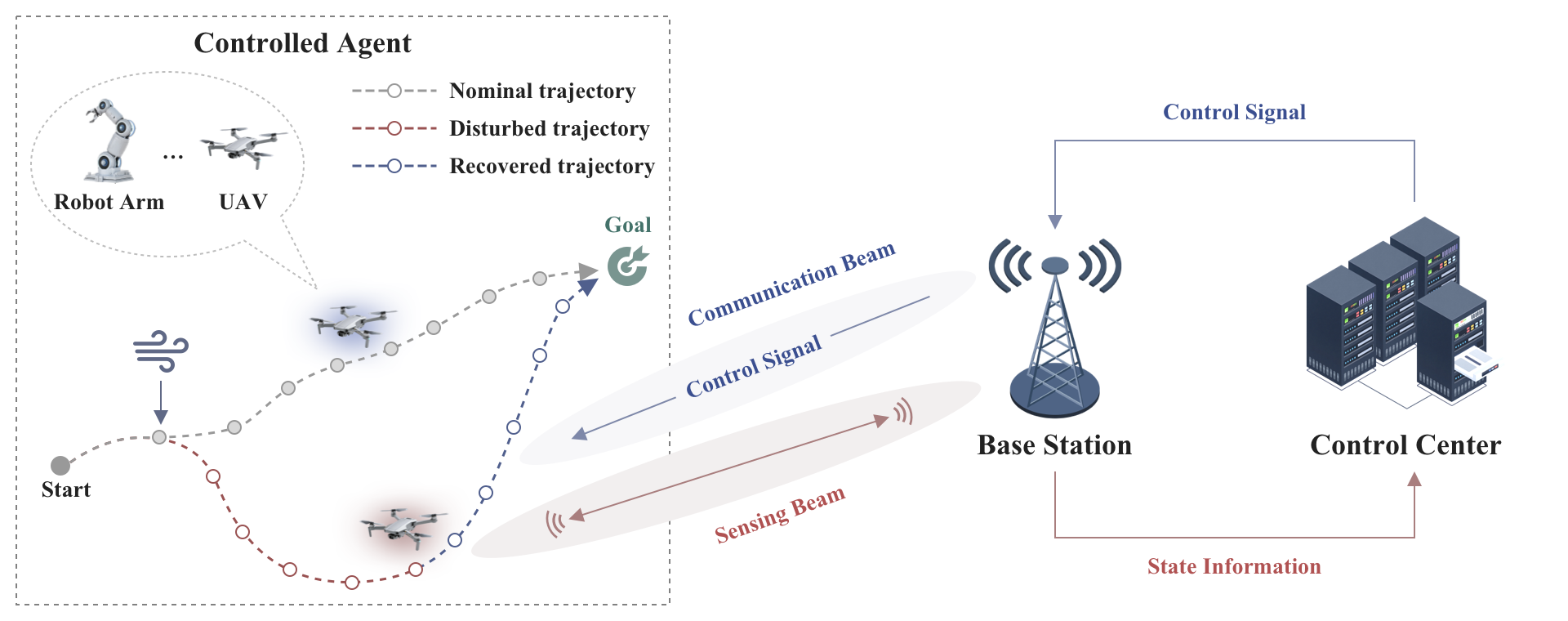}
    \caption{System architecture of the proposed Lyapunov-constrained ISCC framework. 
The control center collects the estimated state information from the BS and generates state-aware control commands according to the controlled agent dynamics and stability requirements. 
The BS then jointly performs downlink communication and radar sensing by transmitting control signals to the controlled agent and receiving sensing echoes for state estimation. 
The left part illustrates the closed-loop recovery process, where the controlled agent deviates from the nominal trajectory under external disturbances and is subsequently steered back toward the desired trajectory through real-time control.}
    \label{fig:system_model}
\end{figure*}
As shown in Fig.~\ref{fig:system_model}, we consider a BS and a controlled agent. The BS is equipped with $N_t$ transmit antennas and $N_r$ receive antennas, and employs a dual-functional downlink waveform with beamforming vector $\mathbf{w}_k$ to simultaneously transmit control commands and sense the state of the controlled agent. The controlled agent may
represent an unmanned aerial vehicle (UAV), a robotic arm, or other cyber-physical system requiring
wireless feedback control.

At each discrete time slot $k$, the BS transmits a downlink signal to deliver control commands. By applying linear transmit beamforming, the transmitted signal is given by
\begin{equation}
    \mathbf{x}_k = \mathbf{w}_k s_k,
\end{equation}
where $\mathbf{w}_k \in \mathbb{C}^{N_t \times 1}$ denotes the transmit beamforming vector, and $s_k$ is the data symbol carrying the control command. The data symbol $s_k$ is an independent and identically distributed (i.i.d.) random variable with normalized power, i.e., $\mathbb{E}[|s_k|^2] = 1$. Furthermore, the wireless channel between the BS and the controlled agent is assumed to be dominated by the line-of-sight (LoS) link.

Throughout this work, the BS is assumed to have perfect
instantaneous CSI, while the aggregate sensing-processing-control
latency is negligible relative to the control sampling period
$T_s$. Accordingly, the stability results characterize a
slot-synchronous baseline and are conditional on these
assumptions. Imperfect-CSI and delay-aware extensions are left
for future work.

\subsection{Communication Model}
At discrete time slot $k$, the received baseband signal at the controlled agent is given by
\begin{equation}
    y_{c,k} = \mathbf{h}_{c,k}^H \mathbf{w}_k s_k + n_{c,k},
\end{equation}
where $\mathbf{h}_{c,k} \in \mathbb{C}^{N_t \times 1}$ denotes the downlink channel vector, and $n_{c,k}$ represents the additive white Gaussian noise (AWGN) at the receiver with variance $\sigma_c^2$.

In the absence of multi-user interference, the received SINR simplifies to
\begin{equation}
    \mathrm{SINR}_k = \frac{|\mathbf{h}_{c,k}^H \mathbf{w}_k|^2}{\sigma_c^2}.
\end{equation}

The instantaneous achievable channel capacity (in bit/s/Hz) of the downlink control link is
\begin{equation}
    C_k = \log_2(1 + \mathrm{SINR}_k).
\end{equation}

Let $R_{\text{th}}$ denote the minimum transmission rate required by
the nominal control link. A packet outage occurs when the instantaneous capacity falls below this threshold, leading to a decoding failure. The corresponding hard outage event is characterized by the indicator function
\begin{equation}
    I_{\text{out},k}(\mathbf{w}_k) = \mathbb{I}\{C_k < R_{\text{th}}\} = \mathbb{I}\{\mathrm{SINR}_k < \gamma\},
\end{equation}
where $\gamma = 2^{R_{\text{th}}} - 1$ is the equivalent SINR decoding threshold.

In short-packet control transmission, the packet decoding outcome is not accurately characterized by an ideal hard-threshold function of the received SINR. Instead, finite-blocklength coding theory shows that the decoding error probability varies smoothly with the received SINR, where the transition region depends on the coding rate, blocklength, and channel dispersion~\cite{Polyanskiy2010FiniteBlocklength,Durisi2016ShortPacket}. Motivated by this finite-blocklength behavior, the communication success probability, i.e., the probability of successful packet delivery, is modeled by a threshold-centered sigmoid function as
\begin{equation}
    p_k(\mathbf{w}_k)
    =
    \frac{1}
    {1+\exp\left[-a\left(\mathrm{SINR}_k-\gamma_{\mathrm{th}}\right)\right]},
\end{equation}
where $\gamma_{\mathrm{th}}$ denotes the SINR transition point associated with the considered short-packet transmission setting, and $a>0$ controls the steepness of the reliability transition. A larger $a$ corresponds to a sharper decoding transition and approaches the conventional hard-threshold outage model, whereas a smaller $a$ captures a wider uncertainty region induced by finite blocklength and practical decoding effects. The corresponding packet outage probability is given by
\begin{equation}
    p_{\mathrm{out},k}(\mathbf{w}_k)
    =
    1-p_k(\mathbf{w}_k).
\end{equation}

\subsection{Control Model}
Since wireless transmissions may fail, let $\alpha_k \in \{0,1\}$ denote the binary packet delivery indicator at time slot $k$. Following the physical-layer communication model, $\alpha_k$ is a Bernoulli random variable with communication success probability $\Pr(\alpha_k = 1) = p_k(\mathbf{w}_k)$ \cite{Schenato2007NCS}.

Assuming negligible processing and over-the-air (OTA) delays, the discrete-time state evolution of the controlled agent is governed by
\begin{equation}
    \mathbf{q}_{k+1}
    =
    \mathbf{A}\mathbf{q}_{k}
    +
    \alpha_k\mathbf{B}\mathbf{u}_{k}
    +
    \boldsymbol{\sigma}_{p,k},
    \label{eq:state_evolution}
\end{equation}
where $\mathbf{q}_k$ denotes the system state relative to the desired operating point. Accordingly, $\|\mathbf{q}_k\|$ measures the system error. The matrices $\mathbf{A}$ and $\mathbf{B}$ are the state transition and input matrices, respectively, and $\boldsymbol{\sigma}_{p,k}$ denotes the physical process noise. The process noise is assumed to be zero-mean and independent across time, with
\begin{equation}
    \mathbb{E}[\boldsymbol{\sigma}_{p,k}]
    =
    \mathbf{0},
    \qquad
    \mathbb{E}[\boldsymbol{\sigma}_{p,k}\boldsymbol{\sigma}_{p,k}^{T}]
    =
    \boldsymbol{\Sigma}_p .
\end{equation}

We adopt a nominal stabilizing LQR controller for an infinite-horizon quadratic cost and account for packet dropouts in the subsequent stability analysis~\cite{Schenato2007NCS,Gatsis_TAC_2014}. The control law is
\begin{equation}
    \mathbf{u}_k
    =
    -\mathbf{K}\mathbf{q}_k,
    \label{eq:lqr_control_structure}
\end{equation}
where $\mathbf{K}$ is the nominal LQR feedback gain.

This nominal controller is adopted to separate the control synthesis from the physical-layer beamforming optimization. In a packet-dropping control system, the optimal feedback gain and the corresponding cost matrix may depend on the communication success probability~\cite{Schenato2007NCS}. In the considered ISCC system, however, this probability is determined by the transmit beamformer through $p_k(\mathbf{w}_k)$. Directly embedding a packet-dependent Riccati recursion into the beamforming design would make both the feedback gain and the Lyapunov matrix functions of the optimization variable $\mathbf{w}_k$, leading to a strongly coupled control-and-communication problem.

To avoid such optimization-dependent control matrices, we use the standard discrete-time algebraic Riccati equation to obtain a fixed stabilizing controller. The nominal LQR gain is computed as
\begin{equation}
    \mathbf{K}
    =
    \left(
    \mathbf{R}
    +
    \mathbf{B}^{T}
    \mathbf{P}
    \mathbf{B}
    \right)^{-1}
    \mathbf{B}^{T}
    \mathbf{P}
    \mathbf{A},
    \label{eq:lqr_gain}
\end{equation}
where $\mathbf{P}\succ\mathbf{0}$ is the solution to
\begin{equation}
    \mathbf{P}
    =
    \mathbf{Q}
    +
    \mathbf{A}^{T}
    \mathbf{P}
    \mathbf{A}
    -
    \mathbf{A}^{T}
    \mathbf{P}
    \mathbf{B}
    \left(
    \mathbf{R}
    +
    \mathbf{B}^{T}
    \mathbf{P}
    \mathbf{B}
    \right)^{-1}
    \mathbf{B}^{T}
    \mathbf{P}
    \mathbf{A}.
    \label{eq:dare}
\end{equation}
Here, $\mathbf{Q}\succeq\mathbf{0}$ and $\mathbf{R}\succ\mathbf{0}$ are the weighting matrices for the system error and the control effort, respectively. The matrix $\mathbf{P}$ defines the quadratic Lyapunov function used in the subsequent stability analysis.

This treatment does not ignore wireless unreliability. Although the feedback gain $\mathbf{K}$ is fixed, the actual control action is still applied through the packet delivery process in \eqref{eq:state_evolution}. Hence, the stabilizing command $\mathbf{u}_k$ affects the plant only when the control packet is successfully delivered. The communication success probability $p_k(\mathbf{w}_k)$ is therefore explicitly retained in the state-dependent Lyapunov constraint derived below. In this way, the controller specifies a stabilizing control direction, while the beamforming optimization adjusts the wireless reliability so that the control action is delivered with sufficient probability for the current system error. This separation keeps the resource allocation problem tractable while preserving the impact of packet dropouts on closed-loop stability.

To evaluate the control performance, the infinite-horizon average LQR cost \cite{Gatsis_TAC_2014} is adopted, given by
\begin{equation}
    J = \limsup_{N\to\infty} \mathbb{E} \left[ \frac{1}{N} \sum_{k=0}^{N-1} \left( \mathbf{q}_k^T\mathbf{Q}\mathbf{q}_k + \mathbf{u}_k^T\mathbf{R}\mathbf{u}_k \right) \right],
    \label{eq:lqr_cost}
\end{equation}
where $N$ is the time horizon. This cost accounts for both the system error and the control energy. A smaller $J$ indicates better closed-loop performance and lower control energy consumption.


%
\section{Problem Formulation}
This section formulates the state-aware Lyapunov-constrained ISCC beamforming problem under ideal and practical sensing.

\subsection{State-Dependent Lyapunov Constraint Under Ideal Sensing}\label{subsec:ideal_sensing}We first consider ideal sensing, where the BS has perfect knowledge of the physical state. Let\begin{equation}V(\mathbf q_k)=\mathbf q_k^{T}\mathbf P\mathbf q_k\end{equation}be a quadratic Lyapunov function, where $\mathbf P\succ\mathbf 0$. We use $V(\mathbf q_k)$ to quantify the instantaneous control risk. Under ideal sensing, the state dynamics in \eqref{eq:state_evolution} become\begin{equation}\mathbf q_{k+1}=\left(\mathbf A-\alpha_k\mathbf B\mathbf K\right)\mathbf q_k+\boldsymbol{\sigma}_{p,k}.\label{eq:ideal_state_dynamics}\end{equation}Using the packet delivery indicator $\alpha_k$ defined in Section II-B, we define\begin{equation}\boldsymbol{\Phi}_{\alpha_k}\triangleq\mathbf A-\alpha_k\mathbf B\mathbf K,\qquad\mathbf A_{\mathrm{cl}}\triangleq\mathbf A-\mathbf B\mathbf K.\end{equation}Substituting \eqref{eq:ideal_state_dynamics} into the Lyapunov function gives\begin{align}V(\mathbf q_{k+1})={}&\mathbf q_k^{T}\boldsymbol{\Phi}_{\alpha_k}^{T}\mathbf P\boldsymbol{\Phi}_{\alpha_k}\mathbf q_k\nonumber\\&+2\mathbf q_k^{T}\boldsymbol{\Phi}_{\alpha_k}^{T}\mathbf P\boldsymbol{\sigma}_{p,k}+\boldsymbol{\sigma}_{p,k}^{T}\mathbf P\boldsymbol{\sigma}_{p,k}.\label{eq:ideal_lyapunov_expansion}\end{align}The process noise is zero mean, independent of $\mathbf q_k$ and $\alpha_k$, and has covariance $\boldsymbol{\Sigma}_p$. Hence, the conditional expectation of the cross term in \eqref{eq:ideal_lyapunov_expansion} is zero, while\begin{equation}\mathbb E\!\left[\boldsymbol{\sigma}_{p,k}^{T}\mathbf P\boldsymbol{\sigma}_{p,k}\right]=\operatorname{Tr}\!\left(\mathbf P\boldsymbol{\Sigma}_p\right).\end{equation}When the packet is delivered, $\alpha_k=1$ and $\boldsymbol{\Phi}_{\alpha_k}=\mathbf A_{\mathrm{cl}}$. When the packet is lost, $\alpha_k=0$ and $\boldsymbol{\Phi}_{\alpha_k}=\mathbf A$. We therefore define\begin{align}L_{\mathrm{cl},k}&\triangleq\mathbf q_k^{T}\mathbf A_{\mathrm{cl}}^{T}\mathbf P\mathbf A_{\mathrm{cl}}\mathbf q_k,\\L_{\mathrm{ol},k}&\triangleq\mathbf q_k^{T}\mathbf A^{T}\mathbf P\mathbf A\mathbf q_k.\end{align}Since $\alpha_k$ is Bernoulli with communication success probability $p_k(\mathbf w_k)$, conditioning on its two outcomes yields\begin{align}\mathbb E\!\left[V(\mathbf q_{k+1})\mid\mathbf q_k\right]={}&p_k(\mathbf w_k)L_{\mathrm{cl},k}+\left[1-p_k(\mathbf w_k)\right]L_{\mathrm{ol},k}\nonumber\\&+\operatorname{Tr}\!\left(\mathbf P\boldsymbol{\Sigma}_p\right).\label{eq:ideal_expected_lyapunov}\end{align}For ideal sensing, we impose the one-step Lyapunov non-increase condition\begin{equation}\mathbb E\!\left[V(\mathbf q_{k+1})\mid\mathbf q_k\right]-V(\mathbf q_k)\leq{} 0.\label{eq:ideal_lyapunov_condition}\end{equation}Substituting \eqref{eq:ideal_expected_lyapunov} into \eqref{eq:ideal_lyapunov_condition} and collecting the terms that contain $p_k(\mathbf w_k)$ gives\begin{align}p_k(\mathbf w_k)\left(L_{\mathrm{ol},k}-L_{\mathrm{cl},k}\right)\geq{}&L_{\mathrm{ol},k}+\operatorname{Tr}\!\left(\mathbf P\boldsymbol{\Sigma}_p\right)\nonumber\\&-V(\mathbf q_k).\label{eq:ideal_reliability_inequality}\end{align}This work considers the case where successful packet delivery reduces the one-step Lyapunov value, i.e., $L_{\mathrm{cl},k}<L_{\mathrm{ol},k}$. Since $L_{\mathrm{ol},k}-L_{\mathrm{cl},k}>0$, \eqref{eq:ideal_reliability_inequality} yields the following minimum communication success probability:\begin{equation}p_k^{\mathrm{req}}\triangleq\frac{L_{\mathrm{ol},k}+\operatorname{Tr}\!\left(\mathbf P\boldsymbol{\Sigma}_p\right)-V(\mathbf q_k)}{L_{\mathrm{ol},k}-L_{\mathrm{cl},k}}.\label{eq:ideal_required_probability}\end{equation}If $p_k^{\mathrm{req}}\leq 0$, the Lyapunov condition is already satisfied and imposes no additional reliability requirement. If $0<p_k^{\mathrm{req}}<1$, it gives a finite SINR requirement. If $p_k^{\mathrm{req}}\geq 1$, the one-step condition cannot be satisfied with finite transmit power under the sigmoid packet-delivery model. This case can occur near the equilibrium when persistent process noise prevents a strict one-step decrease. For $0<p_k^{\mathrm{req}}<1$, inverting the packet-delivery model defined in the communication model gives\begin{equation}\gamma_k^{\mathrm{req}}\triangleq\gamma_{\mathrm{th}}+\frac{1}{a}\ln\!\left(\frac{p_k^{\mathrm{req}}}{1-p_k^{\mathrm{req}}}\right).\label{eq:ideal_required_sinr}\end{equation}Combining this state-dependent threshold with the minimum communication SINR
$\gamma$ gives\begin{equation}\gamma_k^{\min}\triangleq\begin{cases}\gamma, & p_k^{\mathrm{req}}\leq 0,\\[1mm]\max\!\left\{\gamma,\gamma_k^{\mathrm{req}}\right\},&0<p_k^{\mathrm{req}}<1.\end{cases}\label{eq:ideal_effective_sinr}\end{equation}The ideal-sensing beamforming problem is therefore\begin{subequations}\label{prob:ideal_beamforming}\begin{align}\mathcal P_0:\quad\min_{\mathbf w_k}\quad&\|\mathbf w_k\|^2\\\mathrm{s.t.}\quad&|\mathbf h_{c,k}^{H}\mathbf w_k|^2\geq{}\gamma_k^{\min}\sigma_c^2.\end{align}\end{subequations}Assume that $\mathbf h_{c,k}\neq\mathbf 0$, $L_{\mathrm{ol},k}>L_{\mathrm{cl},k}$, and $p_k^{\mathrm{req}}<1$. The objective and constraint in \eqref{prob:ideal_beamforming} are unchanged by a common phase rotation of $\mathbf w_k$. We can therefore choose the phase such that $\mathbf h_{c,k}^{H}\mathbf w_k$ is real and nonnegative. The SINR constraint is then equivalent to\begin{equation}\operatorname{Re}\!\left\{\mathbf h_{c,k}^{H}\mathbf w_k\right\}\geq{}\sigma_c\sqrt{\gamma_k^{\min}},\qquad\operatorname{Im}\!\left\{\mathbf h_{c,k}^{H}\mathbf w_k\right\}=0,\label{eq:ideal_convex_form}\end{equation}which is affine in $\mathbf w_k$. Thus, \eqref{prob:ideal_beamforming} admits an equivalent convex formulation. Moreover, the Cauchy--Schwarz inequality gives\begin{equation}|\mathbf h_{c,k}^{H}\mathbf w_k|^2\leq{}\|\mathbf h_{c,k}\|^2\|\mathbf w_k\|^2.\end{equation}Every feasible beamformer must therefore satisfy\begin{equation}\|\mathbf w_k\|^2\geq{}\frac{\gamma_k^{\min}\sigma_c^2}{\|\mathbf h_{c,k}\|^2}.\end{equation}Equality holds when $\mathbf w_k$ is aligned with $\mathbf h_{c,k}$. Hence, the optimal beamformer is\begin{equation}\mathbf w_k^{\star}=\frac{\sigma_c\sqrt{\gamma_k^{\min}}}{\|\mathbf h_{c,k}\|^2}\mathbf h_{c,k},\label{eq:ideal_closed_form}\end{equation}and the minimum transmit power is\begin{equation}\|\mathbf w_k^{\star}\|^2=\frac{\gamma_k^{\min}\sigma_c^2}{\|\mathbf h_{c,k}\|^2}.\label{eq:ideal_min_power}\end{equation}\subsection{State-Dependent Lyapunov Constraint Under Practical Sensing} In practical systems, the physical state is not directly available; therefore, the BS estimates the controlled agent's kinematic state from the received radar echo. At time slot k, after the dual-functional downlink signal $\mathbf{x}_k=\mathbf{w}_k s_k$ is reflected by the controlled agent, the received baseband echo signal at the BS can be written as \begin{equation} \mathbf{y}_{s,k}=\mathbf{G}_{s,k}\mathbf{x}_k+\mathbf{n}_{s,k}, \end{equation} where $\mathbf{G}_{s,k} \in \mathbb{C}^{N_r \times N_t}$ denotes the target response matrix, and $\mathbf{n}_{s,k}$ denotes the equivalent AWGN vector. Notably, $\mathbf{x}_k$ is not a dedicated radar waveform but a dual-functional signal carrying the real-time control command $s_k$. By modeling the controlled agent as a point sensing target, the response matrix can be written as \begin{equation} \mathbf{G}_{s,k}=\beta\,\mathbf{b}(\theta)\mathbf{a}^H(\theta), \label{eq:Gs_model} \end{equation} where $\beta\in\mathbb{C}$ denotes the complex reflection coefficient. It captures the combined effect of target scattering and propagation loss. The vectors $\mathbf{a}(\theta)\in\mathbb{C}^{N_t\times1}$ and $\mathbf{b}(\theta)\in\mathbb{C}^{N_r\times1}$ are the transmit and receive steering vectors corresponding to the angle $\theta$ of the controlled agent, respectively. To quantify the sensing accuracy for the angle $\theta$, we use the CRB \cite{Liu_TSP_2022} as the performance metric. Assuming $L$ independent snapshots and applying the far-field narrowband approximation ($\dot{\mathbf{b}}(\theta) \approx \mathbf{0}$), the total Fisher information is derived as $\frac{2L|\beta|^2 N_r}{\sigma_s^2} \left|\dot{\mathbf{a}}^H(\theta)\mathbf{w}_k\right|^2$. Consequently, the CRB for angle estimation is \begin{equation} \mathrm{CRB}_{\theta}(\mathbf{w}_k) = \frac{\sigma_s^2}{2L|\beta|^2 N_r \left|\dot{\mathbf{a}}^H(\theta)\mathbf{w}_k\right|^2}, \end{equation} where $\dot{\mathbf{a}}(\theta) = \partial\mathbf{a}(\theta)/\partial\theta$ denotes the spatial derivative of the transmit steering vector. We assume that the sensing system operates in a sufficiently high signal-to-noise ratio (SNR) regime and employs an efficient estimator, such that the angular estimation-error variance is well approximated by the corresponding CRB~\cite{Stoica_Nehorai_1989,Liu_TSP_2022}. Let $\mathbf{e}_k=\mathbf{q}_k-\hat{\mathbf{q}}_k$ denote the conditionally zero-mean state-estimation error. We further assume that the two normalized estimation errors are conditionally uncorrelated and have the same variance. Accordingly, we adopt \begin{equation} \boldsymbol{\Sigma}_{e,k}(\mathbf{w}_k) \triangleq \mathbb{E}\!\left[\mathbf{e}_k\mathbf{e}_k^T\mid\mathbf{w}_k\right] \approx \mathrm{CRB}_{\theta}(\mathbf{w}_k)\mathbf{I}_2. \label{eq:crb_state_error_model} \end{equation} This is an estimator-performance model rather than a claim about an arbitrary estimator. The following stability result is therefore conditional on the stated sensing assumptions. The Lyapunov analysis only uses the conditional error covariance and does not require a specific error distribution. For a non-efficient estimator, the same analysis applies after replacing \eqref{eq:crb_state_error_model} with a calibrated covariance model. According to the nominal LQR control law in \eqref{eq:lqr_control_structure}, the control command is computed based on the estimated state. Combining this control law with the state evolution in \eqref{eq:state_evolution}, the closed-loop state update under practical sensing is given by \begin{equation} \mathbf{q}_{k+1} = \left( \mathbf{A} - \alpha_k \mathbf{B} \mathbf{K} \right) \mathbf{q}_k + \alpha_k \mathbf{B} \mathbf{K} \mathbf{e}_k + \boldsymbol{\sigma}_{p,k}. \label{eq:practical_closed_loop_state} \end{equation} Using $\boldsymbol{\Phi}_{\alpha_k}$ defined in Section III-A, we define \begin{equation} \mathbf{d}_k \triangleq \alpha_k \mathbf{B} \mathbf{K} \mathbf{e}_k + \boldsymbol{\sigma}_{p,k}. \label{eq:Phi_d_def} \end{equation} Then, \eqref{eq:practical_closed_loop_state} can be compactly written as \begin{equation} \mathbf{q}_{k+1} = \boldsymbol{\Phi}_{\alpha_k} \mathbf{q}_k + \mathbf{d}_k. \label{eq:compact_practical_closed_loop} \end{equation} To characterize the one-step stochastic evolution of the closed-loop system, we substitute \eqref{eq:compact_practical_closed_loop} into the quadratic Lyapunov function $V(\mathbf{q}_k)=\mathbf{q}_k^T\mathbf{P}\mathbf{q}_k$. The Lyapunov function value at the next time slot is expanded as \begin{equation} \begin{aligned} V(\mathbf{q}_{k+1}) ={}& \left( \boldsymbol{\Phi}_{\alpha_k}\mathbf{q}_k+\mathbf{d}_k \right)^T \mathbf{P} \left( \boldsymbol{\Phi}_{\alpha_k}\mathbf{q}_k+\mathbf{d}_k \right)\\ ={}& \mathbf{q}_k^T \boldsymbol{\Phi}_{\alpha_k}^T \mathbf{P} \boldsymbol{\Phi}_{\alpha_k} \mathbf{q}_k\\ &+ 2\mathbf{q}_k^T \boldsymbol{\Phi}_{\alpha_k}^T \mathbf{P}\mathbf{d}_k + \mathbf{d}_k^T\mathbf{P}\mathbf{d}_k . \end{aligned} \end{equation} Taking the conditional expectation given $\mathbf{q}_k$ and treating $\mathbf{w}_k$ as fixed, we assume that $\mathbf{e}_k$ and $\boldsymbol{\sigma}_{p,k}$ are conditionally zero mean and mutually independent of the packet-delivery variable $\alpha_k$. Hence, all cross-terms vanish. Define the error weight matrix as \begin{equation} \mathbf{\Omega} \triangleq \mathbf{K}^T\mathbf{B}^T \mathbf{P}\mathbf{B}\mathbf{K}. \end{equation} Using $\alpha_k^2=\alpha_k$, the covariance model in \eqref{eq:crb_state_error_model}, and the definition of $\mathbf{\Omega}$, we obtain \[ \operatorname{Tr}\!\left( \mathbf{\Omega}\boldsymbol{\Sigma}_{e,k}(\mathbf{w}_k) \right) \approx \mathrm{CRB}_{\theta}(\mathbf{w}_k) \operatorname{Tr}(\mathbf{\Omega}). \] The conditional Lyapunov expectation becomes \begin{equation} \begin{aligned} \mathbb{E}\!\left[ V(\mathbf{q}_{k+1})\mid\mathbf{q}_k \right] ={}& p_k(\mathbf{w}_k) \mathbf{q}_k^T \mathbf{A}_{\mathrm{cl}}^T \mathbf{P} \mathbf{A}_{\mathrm{cl}} \mathbf{q}_k\\ &+ \left(1-p_k(\mathbf{w}_k)\right) \mathbf{q}_k^T \mathbf{A}^T\mathbf{P}\mathbf{A} \mathbf{q}_k\\ &+ p_k(\mathbf{w}_k) \mathrm{CRB}_{\theta}(\mathbf{w}_k) \operatorname{Tr}(\mathbf{\Omega})\\ &+ \operatorname{Tr}(\mathbf{P}\boldsymbol{\Sigma}_p), \end{aligned} \label{eq:conditional_lyapunov_practical} \end{equation} where $\operatorname{Tr}(\mathbf{P}\boldsymbol{\Sigma}_p)$ is the beamforming-independent process-noise contribution. Since persistent process and sensing errors may prevent exact convergence to the equilibrium, we impose the additive Lyapunov condition $\mathbb{E}[V(\mathbf{q}_{k+1})\mid\mathbf{q}_k] \leq \rho V(\mathbf{q}_k)+C_0$, where $0<\rho<1$ and $C_0>0$ is a fixed tolerance for the residual effects of these errors. Substituting \eqref{eq:conditional_lyapunov_practical} yields the following beamforming-dependent constraint: \begin{equation} \begin{aligned} & p_k(\mathbf{w}_k) \mathbf{q}_k^T \mathbf{A}_{\mathrm{cl}}^T \mathbf{P} \mathbf{A}_{\mathrm{cl}} \mathbf{q}_k\\ &+ \left(1-p_k(\mathbf{w}_k)\right) \mathbf{q}_k^T \mathbf{A}^T\mathbf{P}\mathbf{A} \mathbf{q}_k\\ &+ p_k(\mathbf{w}_k) \mathrm{CRB}_{\theta}(\mathbf{w}_k) \operatorname{Tr}(\mathbf{\Omega})\\ &+ \operatorname{Tr}(\mathbf{P}\boldsymbol{\Sigma}_p)\\ &\leq \rho\, \mathbf{q}_k^T\mathbf{P}\mathbf{q}_k + C_0 . \end{aligned} \label{eq:joint_lyapunov1} \end{equation} Although a constant contraction factor is mathematically valid, it imposes the same relative Lyapunov contraction requirement at all system-error levels. To require a stronger contraction when the instantaneous system error is large while avoiding an unbounded tightening of the constraint, we define \begin{equation} \rho(\|\mathbf{q}_k\|) = \rho_0-c\tanh(\|\mathbf{q}_k\|), \label{eq:dynamic_rho} \end{equation} where $0<c<\rho_0<1$. The hyperbolic tangent is selected as a simple smooth, monotone, and bounded mapping: it is approximately linear near the equilibrium and gradually saturates for large system errors. Consequently, \[ \rho_0-c < \rho(\|\mathbf{q}_k\|) \leq \rho_0 < 1, \] where $\rho_0$ specifies the near-equilibrium contraction level and $c$ determines the maximum additional tightening. The boundedness result is not specific to $\tanh(\cdot)$; the same proof applies to any nondecreasing saturating function $\varphi(\cdot)$ satisfying $\varphi(0)=0$ and $0\leq\varphi(r)<1$, provided that the resulting per-slot beamforming problem remains feasible. Incorporating the state-dependent contraction factor, the joint Lyapunov constraint is formulated as \begin{equation} \begin{aligned} & p_k(\mathbf{w}_k) \mathbf{q}_k^T \mathbf{A}_{\mathrm{cl}}^T \mathbf{P} \mathbf{A}_{\mathrm{cl}} \mathbf{q}_k\\ &+ \left(1-p_k(\mathbf{w}_k)\right) \mathbf{q}_k^T \mathbf{A}^T\mathbf{P}\mathbf{A} \mathbf{q}_k\\ &+ p_k(\mathbf{w}_k) \mathrm{CRB}_{\theta}(\mathbf{w}_k) \operatorname{Tr}(\mathbf{\Omega})\\ &+ \operatorname{Tr}(\mathbf{P}\boldsymbol{\Sigma}_p)\\ &\leq \rho(\|\mathbf{q}_k\|) \mathbf{q}_k^T\mathbf{P}\mathbf{q}_k + C_0 . \end{aligned} \label{eq:joint_lyapunov} \end{equation} To establish the long-term stability of the closed-loop system under persistent sensing errors and process noise, we next characterize the asymptotic behavior of the system state. \textbf{Theorem 1 (Per-Slot Feasibility and Mean-Square Ultimate Boundedness).} Using $L_{\mathrm{cl},k}$ and $L_{\mathrm{ol},k}$ defined in Section III-A, suppose that $L_{\mathrm{ol},k}\geq L_{\mathrm{cl},k}$. Assume further that there exists a beam direction $\mathbf{v}_k$ satisfying \begin{equation} |\mathbf{h}_{c,k}^H\mathbf{v}_k|>0, \qquad |\dot{\mathbf{a}}^H(\theta)\mathbf{v}_k|>0, \end{equation} and that the fixed controller provides the ideal-link contraction margin \begin{equation} L_{\mathrm{cl},k} + \operatorname{Tr}(\mathbf{P}\boldsymbol{\Sigma}_p) < \rho(\|\mathbf{q}_k\|) \mathbf{q}_k^T\mathbf{P}\mathbf{q}_k + C_0. \label{eq:controller_feasibility} \end{equation} Then, for every such state $\mathbf{q}_k$, there exists a finite-norm beamforming vector $\mathbf{w}_k$ satisfying \eqref{eq:joint_lyapunov}. If this condition holds along the closed-loop trajectory and the corresponding feasible beamformer is applied at every time slot, the closed-loop system is mean-square ultimately bounded, with \begin{equation} \limsup_{k\to\infty} \mathbb{E}\!\left[\|\mathbf{q}_k\|^2\right] \leq \frac{C_0} {\lambda_{\min}(\mathbf{P})(1-\rho_0)}. \end{equation}  \textit{Proof:} See Appendix A. \subsection{Lyapunov-Constrained ISCC Beamforming Optimization Problem}

We then formulate the state-aware beamforming problem. The objective is to minimize the BS transmit power subject to the communication QoS requirement and the state-dependent Lyapunov constraint. The transmit beamforming vector $\mathbf{w}_k \in \mathbb{C}^{N_t \times 1}$ is the sole optimization variable. We define the conditional Lyapunov function value as $\mathcal{L}_k(\mathbf{w}_k) \triangleq \mathbb{E}[V(\mathbf{q}_{k+1}) \mid \mathbf{q}_k]$. The joint ISCC optimization problem at time slot $k$ is formulated as
\begin{equation}
\begin{aligned}
    \mathcal{P}_1: \quad \min_{\mathbf{w}_k} \quad & \|\mathbf{w}_k\|^2 \\
    \mathrm{s.t.} \quad & \mathrm{SINR}_k(\mathbf{w}_k) \ge \gamma, \\
    & \mathcal{L}_k(\mathbf{w}_k) \le \rho(\|\mathbf{q}_k\|) V(\mathbf{q}_k) +C_0.
\end{aligned}
\label{eq:P1}
\end{equation}

Problem $\mathcal{P}_1$ is non-convex because its Lyapunov constraint couples communication reliability with beam-dependent sensing uncertainty.

\section{Proposed Solutions}
\subsection{Problem Transformation and Challenge Analysis}

To reveal the structure of the optimization problem, we reformulate it using SDR. The original optimization variable is the beamforming vector $\mathbf{w}_k$. We define the lifted positive semidefinite matrix~\cite{Luo2010SDR}
\begin{equation}
    \mathbf{W}_k = \mathbf{w}_k \mathbf{w}_k^H.
\end{equation}
This matrix lifting operation introduces a non-convex rank-one constraint
\begin{equation}
    \operatorname{rank}(\mathbf{W}_k) = 1.
\end{equation}

We define the communication channel covariance matrix as
\begin{equation}
    \mathbf{H}_k \triangleq \mathbf{h}_{c,k} \mathbf{h}_{c,k}^H.
\end{equation}
Consequently, the objective of minimizing transmit power is rewritten as a linear trace function with respect to $\mathbf{W}_k$:
\begin{equation}
    \|\mathbf{w}_k\|^2 = \mathrm{Tr}(\mathbf{W}_k).
\end{equation}
Similarly, the SINR constraint for the control link can be rewritten as the affine inequality:
\begin{equation}
    \mathrm{Tr}(\mathbf{H}_k \mathbf{W}_k) \ge \gamma \sigma_c^2.
\end{equation}

Through these algebraic transformations, the objective function and the communication constraint are convexified. The core non-convexity of this problem originates from the state-dependent Lyapunov constraint. Let the sensing array manifold matrix be defined as
\begin{equation}
    \mathbf{J}_\theta \triangleq \dot{\mathbf{a}}(\theta)\dot{\mathbf{a}}^H(\theta).
\end{equation}
To handle the non-differentiable indicator function, the communication success probability is continuously approximated by a sigmoid function based on $\mathbf{W}_k$:
\begin{equation}
    p_k(\mathbf{W}_k) = \frac{1}{1 + \exp\left[-(\tilde{a}\mathrm{Tr}(\mathbf{H}_k \mathbf{W}_k) + \tilde{b})\right]}.
\end{equation}

The sensing CRB can be written exactly as an inverse-trace function:
\begin{equation}
    \mathrm{CRB}_{\theta}(\mathbf{W}_k) = c_{\mathrm{crb}} [\mathrm{Tr}(\mathbf{J}_\theta \mathbf{W}_k)]^{-1},
\end{equation}
where $\tilde{a}$, $\tilde{b}$, and $c_{\mathrm{crb}}$ are aggregated constant coefficients. Specifically, they are defined by the underlying physical parameters as $\tilde{a} \triangleq a/\sigma_c^2$, $\tilde{b} \triangleq -a\gamma_{\mathrm{th}}$, and $c_{\mathrm{crb}} \triangleq \sigma_s^2 / (2L|\beta|^2 N_r)$. 

Substituting these expressions into the original problem yields the following equivalent formulation in terms of $\mathbf{W}_k$:
\begin{equation}
\begin{aligned}
    \min_{\mathbf{W}_k} \quad & \mathrm{Tr}(\mathbf{W}_k) \\
    \mathrm{s.t.} \quad & \mathrm{Tr}(\mathbf{H}_k \mathbf{W}_k) \ge \gamma\sigma_c^2, \\
    & \mathbf{q}_k^T (\mathbf{A}^T\mathbf{P}\mathbf{A})\mathbf{q}_k - \rho(\|\mathbf{q}_k\|)\mathbf{q}_k^T \mathbf{P} \mathbf{q}_k
    + \operatorname{Tr}(\mathbf{P}\boldsymbol{\Sigma}_p) - C_0 \\
    & + p_k(\mathbf{W}_k) \bigg[ \mathbf{q}_k^T(\mathbf{A}_{\mathrm{cl}}^T\mathbf{P}\mathbf{A}_{\mathrm{cl}})\mathbf{q}_k - \mathbf{q}_k^T(\mathbf{A}^T\mathbf{P}\mathbf{A})\mathbf{q}_k \\
    & \qquad \qquad \quad + \mathrm{CRB}_{\theta}(\mathbf{W}_k)\mathrm{Tr}(\mathbf{\Omega}) \bigg] \le 0, \\
    & \mathbf{W}_k \succeq 0, \\
    & \operatorname{rank}(\mathbf{W}_k) = 1.
\end{aligned}
\label{eq:SDR_problem}
\end{equation}

Problem~\eqref{eq:SDR_problem} remains non-convex for two reasons:
the Lyapunov constraint couples the sigmoid communication success
probability with the inverse-trace sensing CRB, and
$\mathbf{W}_k$ is subject to a rank-one constraint. We therefore
relax the rank-one constraint and use SCA to replace the coupled
nonlinear constraint with a locally consistent affine
surrogate~\cite{Scutari_TSP_2017}.

\subsection{Local Linearization and Tractable SCA Formulation}

To streamline the differentiation process, we define the constant matrix term at time slot $k$ as $\Delta_k \triangleq \mathbf{q}_k^T (\mathbf{A}_{\mathrm{cl}}^T \mathbf{P} \mathbf{A}_{\mathrm{cl}}) \mathbf{q}_k - \mathbf{q}_k^T (\mathbf{A}^T \mathbf{P} \mathbf{A}) \mathbf{q}_k$. Then, the original non-convex Lyapunov constraint $F(\mathbf{W}_k) \le 0$ is reformulated as
\begin{equation}
\begin{aligned}
    F(\mathbf{W}_k) \triangleq&~ \mathbf{q}_k^T (\mathbf{A}^T \mathbf{P} \mathbf{A}) \mathbf{q}_k - \rho(\|\mathbf{q}_k\|) \mathbf{q}_k^T \mathbf{P} \mathbf{q}_k
    + \operatorname{Tr}(\mathbf{P}\boldsymbol{\Sigma}_p) - C_0 \\
    & + p_k(\mathbf{W}_k) \left[ \Delta_k + \mathrm{CRB}_{\theta}(\mathbf{W}_k) \mathrm{Tr}(\mathbf{\Omega}) \right] \le 0.
\end{aligned}
\end{equation}

In the SCA framework, we handle $F(\mathbf{W}_k)$ using a
locally consistent affine approximation constructed at the
current iterate $\mathbf{W}_k^{(t)}$. Because
$F(\mathbf{W}_k)$ contains the product of a sigmoid function
and an inverse-trace sensing term, it is generally neither
globally convex nor globally concave. Therefore, its first-order
Taylor approximation is not asserted to be a global majorizer.
Instead, it is used as a local SCA surrogate:
\begin{equation}
    \widehat{F}_t(\mathbf{W}_k) \triangleq F(\mathbf{W}_k^{(t)}) + \mathrm{Tr}\left( \nabla_{\mathbf{W}_k} F(\mathbf{W}_k^{(t)}) (\mathbf{W}_k - \mathbf{W}_k^{(t)}) \right).
\end{equation}

By construction, the affine surrogate satisfies the local
value- and gradient-consistency conditions
\begin{equation}
\widehat{F}_t(\mathbf{W}_k^{(t)})
=
F(\mathbf{W}_k^{(t)}),
\qquad
\nabla_{\mathbf{W}_k}
\widehat{F}_t(\mathbf{W}_k^{(t)})
=
\nabla_{\mathbf{W}_k}
F(\mathbf{W}_k^{(t)}).
\label{eq:sca_consistency}
\end{equation}
We assume that the SCA iterates remain bounded and satisfy
$\operatorname{Tr}(\mathbf{J}_\theta\mathbf{W}_k)
\geq \epsilon_s>0$, so that the inverse-trace sensing term and
its gradient remain well defined. Accordingly, the proposed procedure relies on the standard local SCA conditions rather than on the global majorization property required by majorization--minimization (MM). Under the standard regularity and
step-control conditions~\cite{Scutari_TSP_2017}, the resulting
iterative procedure converges to a stationary solution; neither
global convergence nor global optimality is claimed.

Since $\mathbf{H}_k, \mathbf{J}_\theta$, and $\mathbf{W}_k$ are Hermitian matrices, this approximation is real-valued and strictly affine with respect to $\mathbf{W}_k$. 

Applying the product and chain rules of matrix calculus, the gradient $\nabla_{\mathbf{W}_k} F(\mathbf{W}_k)$ is analytically derived in closed form as
\begin{equation}
\begin{aligned}
\nabla_{\mathbf{W}_k} F(\mathbf{W}_k)
=&~ \tilde{a} p_k(\mathbf{W}_k)
\bigl(1-p_k(\mathbf{W}_k)\bigr)
\\
&\times
\Bigg[
\Delta_k
+
\frac{
c_{\mathrm{crb}}\mathrm{Tr}(\mathbf{\Omega})
}{
\mathrm{Tr}(\mathbf{J}_\theta \mathbf{W}_k)
}
\Bigg]
\mathbf{H}_k
\\
&-
p_k(\mathbf{W}_k)
\frac{
c_{\mathrm{crb}}\mathrm{Tr}(\mathbf{\Omega})
}{
\bigl[\mathrm{Tr}(\mathbf{J}_\theta \mathbf{W}_k)\bigr]^2
}
\mathbf{J}_\theta .
\end{aligned}
\label{eq:gradient_split}
\end{equation}

By replacing the original non-convex Lyapunov constraint with $\widehat{F}_t(\mathbf{W}_k) \le 0$ and temporarily relaxing the rank-one constraint, the convex subproblem at the $t$-th iteration is formulated as
\begin{equation}
\begin{aligned}
    \min_{\mathbf{W}_k} \quad & \mathrm{Tr}(\mathbf{W}_k) \\
    \mathrm{s.t.} \quad & \mathrm{Tr}(\mathbf{H}_k \mathbf{W}_k) \ge \gamma \sigma_c^2, \\
    & \widehat{F}_t(\mathbf{W}_k) \le 0, \\
    & \mathbf{W}_k \succeq 0.
\end{aligned}
\label{eq:SCA_subproblem}
\end{equation}

For completeness, we next clarify the tightness of the rank
relaxation in \eqref{eq:SCA_subproblem}. For a fixed SCA
iteration $t$, $\widehat{F}_t(\mathbf{W}_k)$ is affine in
$\mathbf{W}_k$. Hence, \eqref{eq:SCA_subproblem} is a complex
SDP with one Hermitian positive semidefinite
matrix variable and two scalar affine constraints, namely, the
communication constraint and the linearized Lyapunov constraint.

Assume that the primal and dual problems associated with
\eqref{eq:SCA_subproblem} are solvable. A sufficient condition
is the existence of a strictly feasible matrix
$\widetilde{\mathbf{W}}_k \succ \mathbf{0}$ satisfying
\begin{equation}
\operatorname{Tr}
\left(\mathbf{H}_k\widetilde{\mathbf{W}}_k\right)
>
\gamma \sigma_c^2,
\qquad
\widehat{F}_t
\left(\widetilde{\mathbf{W}}_k\right)<0.
\label{eq:sdr_slater_condition}
\end{equation}
Under this condition, the complex SDP rank-reduction result
in~\cite{Huang_TSP_2010} guarantees the existence of an optimal
solution $\overline{\mathbf{W}}_k^{(t)}$ satisfying
\begin{equation}
\operatorname{rank}^{2}
\left(\overline{\mathbf{W}}_k^{(t)}\right)
\leq 2,
\label{eq:sdr_rank_bound}
\end{equation}
where the right-hand side is the number of scalar affine
constraints in \eqref{eq:SCA_subproblem}. Moreover, since
$\gamma \sigma_c^2>0$, the communication constraint
excludes the zero matrix, and therefore
$\operatorname{rank}
(\overline{\mathbf{W}}_k^{(t)})\geq1$. Because the rank is an
integer, \eqref{eq:sdr_rank_bound} consequently implies
\begin{equation}
\operatorname{rank}
\left(\overline{\mathbf{W}}_k^{(t)}\right)=1.
\end{equation}
Thus, whenever the fixed-iteration SDP subproblem is solvable,
there exists an optimal beamforming vector
$\overline{\mathbf{w}}_k^{(t)}$ such that
\begin{equation}
\overline{\mathbf{W}}_k^{(t)}
=
\overline{\mathbf{w}}_k^{(t)}
\overline{\mathbf{w}}_k^{(t)H},
\end{equation}
and dropping the rank-one constraint introduces no relaxation
gap at the SCA-subproblem level. This result does not establish
global optimality for the original non-convex problem; the
overall solution remains dependent on the local SCA
approximation.

The overall iterative procedure for solving the joint ISCC beamforming design is summarized in Algorithm 1.

\begin{algorithm}[t]
\caption{Proposed SDR--SCA Beamforming Algorithm}
\label{alg:SCA}
\begin{algorithmic}[1]
\Require System state $\mathbf q_k$, channel and sensing parameters, tolerances $\epsilon$ and $\epsilon_{\mathrm{rank}}$, backtracking factor $\beta\in(0,1)$, maximum iteration number $T_{\max}$, and randomization number $N_{\mathrm{rand}}$
\Ensure Feasible beamforming vector $\mathbf w_k$
\State Initialize $t=0$ and a feasible $\mathbf W_k^{(0)}$ satisfying all constraints of the original problem
\Repeat
    \State Evaluate $p_k(\mathbf W_k^{(t)})$, $\mathrm{CRB}_{\theta}(\mathbf W_k^{(t)})$, and $\nabla_{\mathbf W_k}F(\mathbf W_k^{(t)})$
    \State Construct the affine surrogate $\widehat F_t(\mathbf W_k)$
    \State Solve \eqref{eq:SCA_subproblem} to obtain $\widetilde{\mathbf W}_k^{(t+1)}$
    \State Set $\mathbf D_k^{(t)}=\widetilde{\mathbf W}_k^{(t+1)}-\mathbf W_k^{(t)}$
    \State Choose the largest $\eta_t=\beta^m$, $m=0,1,\ldots$, such that $\mathbf W_k^{(t)}+\eta_t\mathbf D_k^{(t)}$ satisfies the original constraints
    \State Update $\mathbf W_k^{(t+1)}=\mathbf W_k^{(t)}+\eta_t\mathbf D_k^{(t)}$
    \State $t\leftarrow t+1$
\Until{$\|\mathbf W_k^{(t)}-\mathbf W_k^{(t-1)}\|_F/\max\{1,\|\mathbf W_k^{(t-1)}\|_F\}\leq\epsilon$ or $t\geq T_{\max}$}
\State $r_{\mathrm{rank}}\leftarrow[\operatorname{Tr}(\mathbf W_k^{(t)})-\lambda_{\max}(\mathbf W_k^{(t)})]/\operatorname{Tr}(\mathbf W_k^{(t)})$
\If{$r_{\mathrm{rank}}\leq\epsilon_{\mathrm{rank}}$}
    \State $\mathbf w_k\leftarrow\sqrt{\lambda_{\max}(\mathbf W_k^{(t)})}\,\mathbf u_{\max}(\mathbf W_k^{(t)})$
\Else
    \State Generate $N_{\mathrm{rand}}$ candidates from $\mathcal{CN}(\mathbf 0,\mathbf W_k^{(t)})$
    \State Scale each candidate and test it against the original constraints
    \State Select the minimum-power feasible candidate as $\mathbf w_k$
\EndIf
\State \textbf{return} $\mathbf w_k$
\end{algorithmic}
\end{algorithm}

Although the rank-reduction result guarantees the existence of a rank-one optimum for every solvable SCA subproblem, a numerical SDP solver may return a higher-rank solution because of non-uniqueness and finite solver tolerance. When necessary, we recover a rank-one beamformer using Gaussian randomization or dominant-eigenvector extraction~\cite{Sidiropoulos2006Multicasting}. Any resulting power difference is therefore a numerical recovery loss rather than an SDR relaxation gap.

\section{Simulation Results}
This section evaluates how the proposed state-aware design
adapts transmit power, communication reliability, and sensing
accuracy to variations in the physical state and external
disturbances. We further examine closed-loop recovery, ISCC trade-offs, and computational performance to evaluate the role of state-aware
beamforming in supporting physical tasks under external disturbances.

\subsection{Simulation Setup and Benchmark Schemes}

Unless otherwise specified, the simulations use a sub-6 GHz setting inspired by the 3GPP TR 38.901 urban-microcell scenario. The carrier frequency is $f_c=3.5~\mathrm{GHz}$, the system bandwidth is $B=20~\mathrm{MHz}$, the subcarrier spacing is $\Delta f=30~\mathrm{kHz}$, and the antenna spacing is $d=\lambda/2$, where $\lambda$ is the carrier wavelength~\cite{3GPP38901}. The control period is $T_s=1~\mathrm{ms}$, and each control command is transmitted as a short packet.

The BS employs $N_t=4$ transmit antennas and $N_r=4$ receive antennas. The communication and sensing noise powers are $\sigma_c^2=\sigma_s^2=10^{-3}$. The communication parameters are $\gamma=5$, $\gamma_{\mathrm{th}}=4.5$, and $a=0.4$, while the sensing parameters are $L=10$ and $|\beta|^2=1$. These representative parameters provide a feasible nominal operating point and are not independently optimized.

Since communication and sensing involve the same controlled agent, we use a common direction $\theta=20^\circ$ and normalize the LoS channel as $\mathbf{h}_{c,k}=\mathbf{a}(\theta)$. Accordingly,
\begin{equation}
\mathbf{H}_k=\mathbf{a}(\theta)\mathbf{a}^{H}(\theta),
\qquad
\mathbf{J}_{\theta}=\dot{\mathbf{a}}(\theta)\dot{\mathbf{a}}^{H}(\theta).
\end{equation}
Although the communication and sensing links share the same direction, their beam dependencies are characterized by $\mathbf{a}(\theta)$ and $\dot{\mathbf{a}}(\theta)$, respectively. The matrices $\mathbf{H}_k$ and $\mathbf{J}_{\theta}$, together with the aggregated coefficients $\tilde a$, $\tilde b$, and $c_{\mathrm{crb}}$, are derived from the stated physical parameters and are not independently tuned.

The plant dynamics and the LQR weights are set as
\[
\begin{gathered}
\mathbf{A}=
\begin{bmatrix}
1.08 & 0.10\\
0 & 1.05
\end{bmatrix},
\qquad
\mathbf{B}=
\begin{bmatrix}
0\\
1
\end{bmatrix},\\[-1mm]
\mathbf{Q}=\mathbf{I}_2,
\qquad
\mathbf{R}=0.2.
\end{gathered}
\]
The pair $(\mathbf{A},\mathbf{B})$ is controllable, and $\mathbf{A}$ represents a mildly unstable second-order task. The matrices $\mathbf{P}$, $\mathbf{K}$, $\mathbf{A}_{\mathrm{cl}}$, and $\mathbf{\Omega}$ are determined by the stated plant and LQR parameters rather than independently selected. For the normalized two-dimensional state, we use
\begin{equation}
\boldsymbol{\Sigma}_{e,k}(\mathbf{w}_k)\approx
\mathrm{CRB}_{\theta}(\mathbf{w}_k)\mathbf{I}_2.
\end{equation}

We set $\rho_0=0.98$ and $c=0.06$, which gives $\rho(\|\mathbf{q}_k\|)\in(0.92,0.98]$. This range provides stronger contraction when the system error is large while retaining per-slot feasibility in the tested setting. The SDR-SCA algorithm uses $T_{\max}=20$, $\epsilon=10^{-3}$, and $N_{\mathrm{rand}}=200$. Beyond the nominal operating point, the simulations vary the antenna configuration, sensing-noise variance, and process-noise variance to evaluate the sensitivity of the proposed design.

We consider two baseline schemes. The static ISAC baseline employs a fixed beamformer designed under predefined communication and sensing requirements. The control-aware baseline adapts the communication resources using the same state-dependent Lyapunov contraction factor as the proposed method, but excludes the beam-dependent sensing-error term $\mathrm{CRB}_{\theta}(\mathbf{w}_k)$ from the Lyapunov constraint. All schemes are evaluated using the same plant dynamics, controller, channel and sensing models, antenna configuration, noise levels, initial state, and common random-number streams for packet delivery, process noise, and sensing errors.

\begin{figure}[!t]
    \centering
    \includegraphics[width=0.95\columnwidth]{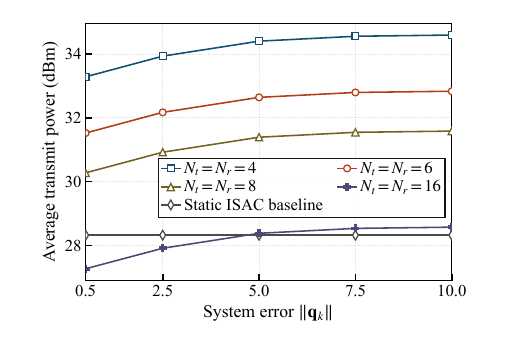}
    \caption{Average transmit power versus the system error $\|\mathbf{q}_k\|$ under different antenna configurations. The gray dashed line denotes the static ISAC baseline. The proposed method increases the transmit power as the system error grows, thereby providing stronger communication and sensing support when the physical-state deviation is large. In contrast, the static ISAC baseline is designed under fixed communication and sensing requirements, which may lead to unnecessary power consumption when the system error is small and insufficient support when the system error becomes large.}
    \label{fig:qk_power_nt_nr_static}
\end{figure}

Fig.~\ref{fig:qk_power_nt_nr_static} shows that the required transmit power increases with the system error and decreases as the antenna array grows. A larger array provides higher communication beamforming gain and better sensing spatial focusing, thereby reducing the power required to satisfy the joint constraint. Unlike the static ISAC baseline, the proposed design adapts its transmit power to the instantaneous system error, avoiding excessive resource allocation at low-error states and insufficient support at high-error states.
\subsection{State-Aware Resource Adaptation and Recovery}

This subsection evaluates how the BS adapts its wireless resources to physical-state evolution. We examine this behavior through the Lyapunov function, system error, achieved SINR, and LQR cost.

\begin{figure}[!t]
    \centering
    \includegraphics[width=\columnwidth]{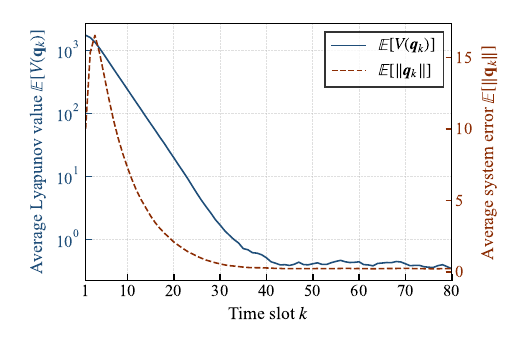}
    \caption{Closed-loop recovery under the proposed Lyapunov-constrained beamforming design. The left axis shows the ensemble-averaged Lyapunov function $\mathbb{E}[V(\mathbf{q}_k)]$, which quantifies the instantaneous control risk, while the right axis shows the ensemble-averaged system error $\mathbb{E}[\|\mathbf{q}_k\|]$, which measures the deviation of the physical state from the desired state. Their large initial values correspond to a strongly disturbed physical state. As the BS adapts its beamforming over successive time slots, both quantities decrease and eventually remain within a bounded operating region. Their joint decrease indicates that the allocated wireless resources provide sufficient packet-delivery reliability and sensing accuracy to reduce the physical-state deviation. The remaining steady-state error is caused by persistent packet losses, sensing errors, and process noise.}
    \label{fig:convergence}
\end{figure}

\begin{figure}[!t]
    \centering
    \includegraphics[width=\columnwidth]{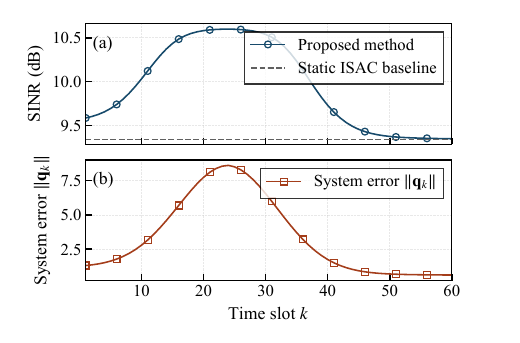}
    \caption{State-dependent SINR adaptation under a time-varying system-error profile. The upper panel shows the achieved SINR of the proposed Lyapunov-constrained ISCC design and the fixed SINR of the static ISAC baseline, while the lower panel shows the corresponding system error $\|\mathbf{q}_k\|$. When the physical state moves farther from the desired state, the system error increases and the state-dependent Lyapunov constraint becomes tighter. The BS consequently increases the achieved SINR to improve packet-delivery reliability. As the system error decreases, the required SINR is relaxed to avoid unnecessary resource consumption. In contrast, the static ISAC baseline maintains the same SINR level and therefore cannot adjust its communication support to the physical-state evolution.}
    \label{fig:dynamic_sinr_system_error}
\end{figure}

\begin{figure}[!t]
    \centering
    \includegraphics[width=\columnwidth]{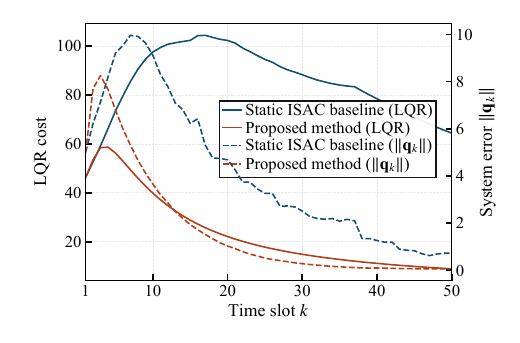}
    \caption{Dynamic trajectories of the LQR cost and the system error $\|\mathbf{q}_k\|$ over 50 time slots. The LQR cost measures the combined effect of physical-state deviation and control effort, while $\|\mathbf{q}_k\|$ directly measures the deviation from the desired physical state. Following the initial disturbance, the proposed Lyapunov-constrained ISCC design produces a lower peak LQR cost and faster system-error decay than the static ISAC baseline. This result indicates that providing stronger wireless support during high-error periods limits the transient performance loss and accelerates the return of the physical task to a bounded operating region.}
    \label{fig:lqr_compare}
\end{figure}

Figs.~\ref{fig:convergence}--\ref{fig:lqr_compare} jointly illustrate the recovery process and its underlying wireless mechanism. Fig.~\ref{fig:convergence} shows that the physical task recovers from the disturbed state, while Fig.~\ref{fig:dynamic_sinr_system_error} shows that this recovery is supported by adapting the communication reliability to the instantaneous system error. The resulting reduction in the LQR cost and system-error decay time in Fig.~\ref{fig:lqr_compare} demonstrates the performance benefit of state-aware wireless resource allocation over fixed resource provisioning.

We further evaluate the proposed method against the control-aware baseline under different sensing-noise levels.

\begin{figure*}[!t]
    \centering
    \includegraphics[width=\textwidth]{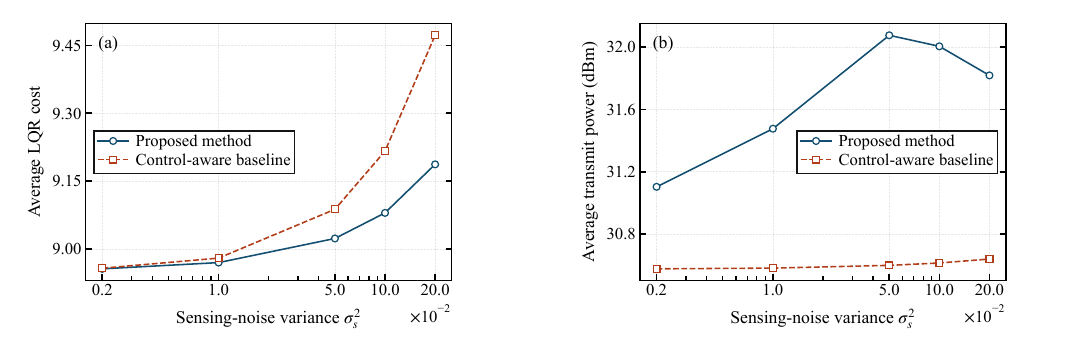}
    \caption{Comparison between the proposed method and the control-aware baseline under different sensing-noise variance levels. (a) Average LQR cost versus the sensing-noise variance $\sigma_s^2$. The two methods achieve comparable LQR costs at low sensing-noise levels, whereas the proposed method achieves a lower LQR cost as $\sigma_s^2$ increases by explicitly incorporating the beam-dependent sensing uncertainty into the state-dependent Lyapunov constraint. (b) Average transmit power versus $\sigma_s^2$. The proposed method generally requires higher transmit power because the shared beam jointly supports packet delivery and state sensing. The non-monotonic power variation results from the joint adjustment of the beamforming power and direction as the optimizer rebalances communication gain and sensing information. Overall, the comparison characterizes the trade-off between physical-task performance and transmit-power expenditure under sensing uncertainty.}
    \label{fig:control_aware_comparison}
\end{figure*}

\begin{figure}[!t]
    \centering
    \includegraphics[width=0.95\columnwidth]{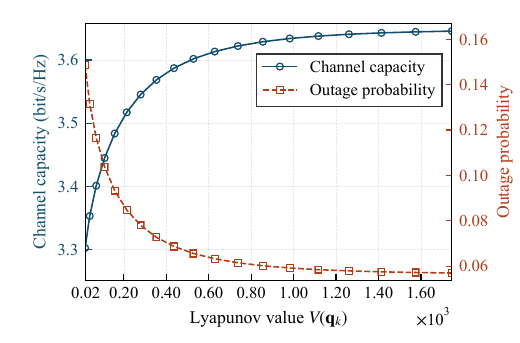}
    \caption{Channel capacity and outage probability versus the instantaneous control risk quantified by the Lyapunov function value $V(\mathbf{q}_k)$. The left axis shows the channel capacity, while the right axis shows the outage probability. As $V(\mathbf{q})$ increases, the proposed Lyapunov-constrained ISCC design allocates more resources to the communication link, leading to a higher channel capacity and a lower outage probability. This result indicates that the communication reliability is enhanced when the physical task exhibits a larger system error.}
    \label{fig:V_capacity_pout}
\end{figure}

\begin{figure}[!t]
    \centering
    \includegraphics[width=0.95\columnwidth]{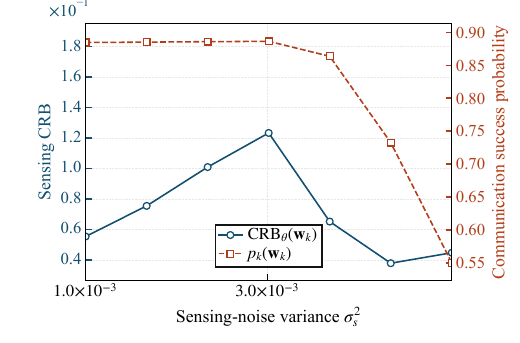}
    \caption{Achieved sensing CRB and communication success probability versus the sensing-noise variance $\sigma_s^2$. The non-monotonic CRB results from the joint adaptation of the power and direction of the shared beam. At low-to-moderate sensing-noise levels, the CRB increases because the sensing-noise degradation dominates. As $\sigma_s^2$ further increases, the Lyapunov-constrained beamforming design allocates more beamforming gain toward the sensing direction to mitigate the beam-dependent estimation uncertainty, thereby reducing the achieved CRB at the expense of the effective communication gain and packet-delivery success probability.}
    \label{fig:crb_psucc_tradeoff}
\end{figure}

\begin{figure}[!t]
    \centering
    \includegraphics[width=0.93\columnwidth]{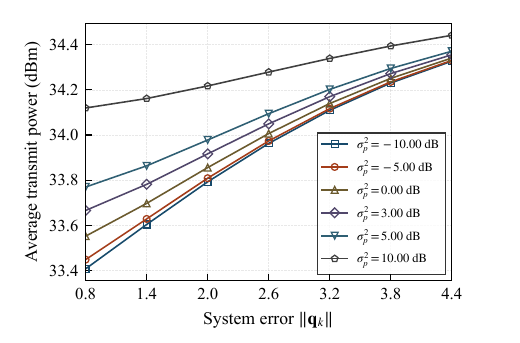}
    \caption{Average transmit power versus the system error $\|\mathbf{q}_k\|$ under different process-noise variance levels. The process-noise variance is represented in dB as $10\log_{10}(\sigma_p^2/\sigma_0^2)$, where the reference variance is $\sigma_0^2=1$. For a given system error, a larger process-noise variance requires higher transmit power to satisfy the stability-related requirement. Under each process-noise level, the transmit power also increases with $\|\mathbf{q}_k\|$, showing that the proposed Lyapunov-constrained ISCC design adapts to both the instantaneous system error and the background disturbance intensity.}
    \label{fig:process_noise_power}
\end{figure}

Fig.~\ref{fig:control_aware_comparison} shows that the two methods achieve comparable LQR costs at low sensing-noise levels, while the proposed method achieves a lower LQR cost as the sensing uncertainty increases. This improvement is obtained at the expense of higher average transmit power, demonstrating a trade-off between physical-task performance and wireless resource consumption.

\begin{figure}[!t]
    \centering
    \includegraphics[width=0.95\columnwidth]{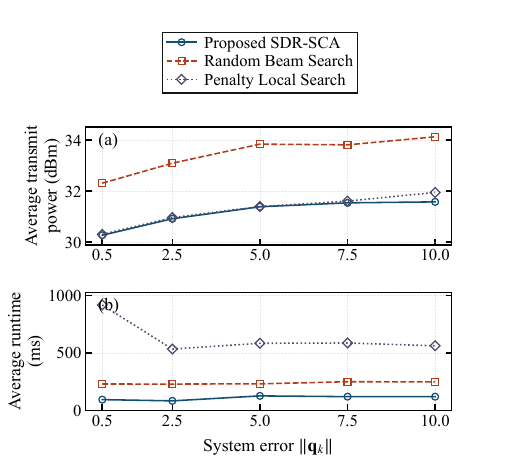}
    \caption{Comparison of different non-convex beamforming solvers under the same SINR requirement and state-dependent Lyapunov constraint. The adopted SDR-SCA solver is compared with random beam search and penalty-based local search. For all methods, the final beamformer is evaluated using the original constraints, and infeasible trials are excluded from the transmit-power averaging. 
    (a) Transmit power versus the system error \(\|\mathbf{q}_k\|\). The SDR-SCA solver requires lower transmit power than the two baseline methods over the considered system-error range. 
    (b) Average runtime versus the system error $\|\mathbf{q}_k\|$. The proposed SDR-SCA solver achieves the lowest average runtime among the three methods over the tested system-error range, while also requiring less transmit power to satisfy the original constraints.}
    \label{fig:solver_power_runtime}
\end{figure}

\begin{figure}[!t]
    \centering
    \includegraphics[width=0.95\columnwidth]{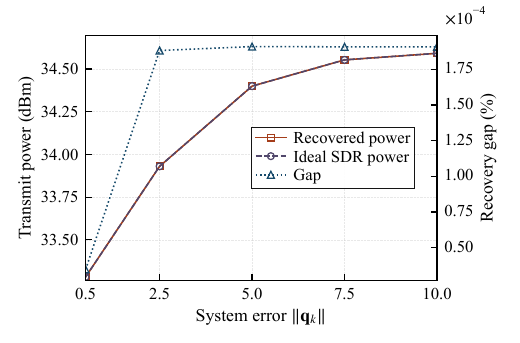}
    \caption{Recovered rank-one beamforming power and the
    relaxed-SDP optimal power versus the system error
    \(\|\mathbf{q}_k\|\). The recovered beamforming power closely
    follows the relaxed-SDP optimum over the considered range. The
    small difference between the two curves quantifies the empirical
    loss introduced by the adopted numerical rank-one recovery
    procedure and solver tolerance; it is not used to establish SDR
    tightness, which is addressed analytically by the rank-reduction
result above.}
    \label{fig:rank1_gap}
\end{figure}

\subsection{Trade-offs in ISCC}

The communication, sensing, and control functions in the proposed Lyapunov-constrained ISCC framework are coupled through shared transmit resources and state-aware beamforming. Improving communication reliability requires additional beamforming gain or transmit power, while maintaining sensing accuracy under degraded sensing conditions also consumes wireless resources. Since both factors affect closed-loop control performance, the ISCC design needs to balance these functions rather than optimizing any single metric alone. The following results examine this coupling from three aspects: instantaneous control risk, sensing degradation, and process-noise intensity.

Fig.~\ref{fig:V_capacity_pout} shows that a larger Lyapunov function value tightens the state-dependent Lyapunov constraint and increases the communication beamforming gain, resulting in higher channel capacity and lower outage probability. Fig.~\ref{fig:crb_psucc_tradeoff} shows that, as the sensing-noise variance increases, the shared beam is rebalanced between sensing and communication, leading to a non-monotonic sensing CRB and a reduction in the communication success probability. Fig.~\ref{fig:process_noise_power} further shows that stronger process noise reduces the available Lyapunov contraction margin and therefore increases the transmit power required at a given system-error level. Taken together, these results characterize the communication--sensing--power trade-off in the considered shared-beamforming design: the BS balances communication reliability and sensing accuracy while adapting its transmit power to the instantaneous system error and disturbance intensity.

\subsection{Algorithm Convergence and Computational Complexity}

In addition to the closed-loop performance, we evaluate the
computational cost of the proposed SDR-SCA solver. According
to Algorithm~\ref{alg:SCA}, the algorithm repeatedly solves
the convex SDP subproblem in \eqref{eq:SCA_subproblem} and
then performs numerical rank-one recovery.

At each outer SCA iteration, the optimization variable is the
Hermitian matrix
$\mathbf{W}_k\in\mathbb{C}^{N_t\times N_t}$. The subproblem
contains one positive-semidefinite cone constraint and two scalar
affine constraints, corresponding to the SINR requirement and
the locally approximated Lyapunov constraint. In our implementation,
this convex subproblem is solved using the splitting conic solver (SCS), a first-order method for
conic optimization. Let $I_{\mathrm{scs}}$ denote the number of
internal SCS iterations required to solve one SDP subproblem.
The dominant operation in each SCS iteration is the projection
onto the positive-semidefinite cone, which requires an eigenvalue
decomposition with complexity $\mathcal{O}(N_t^3)$. Therefore,
the computational cost of one SDP subproblem is approximately
\begin{equation}
    \mathcal{O}\!\left(I_{\mathrm{scs}}N_t^3\right).
    \label{eq:sdp_complexity}
\end{equation}

Let $T_{\mathrm{sca}}$ denote the number of outer SCA iterations.
The computational complexity of the complete SCA stage is then
\begin{equation}
    \mathcal{O}\!\left(
    T_{\mathrm{sca}}I_{\mathrm{scs}}N_t^3
    \right).
    \label{eq:sca_complexity}
\end{equation}

After solving the relaxed SDP, an eigenvalue decomposition of
$\mathbf{W}_k$ is performed at a cost of $\mathcal{O}(N_t^3)$.
If the numerical solution is rank one, the beamforming vector is
recovered directly from its principal eigenpair. Otherwise,
$N_{\mathrm{rand}}$ candidate beamformers are generated and
evaluated using the original constraints. Since each candidate
evaluation requires $\mathcal{O}(N_t^2)$ operations, the numerical
recovery stage has complexity
\begin{equation}
    \mathcal{O}\!\left(
    N_t^3+N_{\mathrm{rand}}N_t^2
    \right).
    \label{eq:recovery_complexity}
\end{equation}

Consequently, the implementation-level computational complexity
of the proposed solver is
\begin{equation}
    \mathcal{O}\!\left(
    T_{\mathrm{sca}}I_{\mathrm{scs}}N_t^3
    +N_t^3+N_{\mathrm{rand}}N_t^2
    \right).
    \label{eq:overall_complexity}
\end{equation}

In the simulations, we set $T_{\max}=20$ and $N_{\mathrm{rand}}=200$. Since the internal SCS iteration count depends on the problem conditioning and stopping tolerance, Fig.~\ref{fig:solver_power_runtime} reports the measured wall-clock runtime together with the transmit power. Over the tested system-error range, the proposed SDR-SCA solver achieves both lower transmit power and lower average runtime than the two numerical baselines. Fig.~\ref{fig:rank1_gap} further shows that the recovered rank-one beamforming power closely follows the relaxed-SDP optimum, indicating only a small empirical recovery loss.

\section{Conclusion}
\label{sec:conclusion}

Within the considered single-BS, single-agent setting, this paper
developed a Lyapunov-constrained ISCC beamforming framework that adapts communication and sensing resources to the instantaneous physical state. The proposed constraint captures packet losses, sensing uncertainty, and process noise, and establishes mean-square ultimate boundedness of the closed-loop state under the stated assumptions. The ideal-sensing problem admits a closed-form solution, whereas the practical-sensing problem is addressed by an SDR--SCA algorithm with tight per-iteration SDR. Simulations demonstrate state-aware resource adaptation and characterize the trade-off between physical-task performance and transmit-power expenditure under sensing uncertainty and process noise. Future work will consider multiple agents,
imperfect CSI, and communication delays.

\appendices
\section{Proof of Theorem 1}
\label{app:theorem1}
For $\mathbf{w}_k=t\mathbf{v}_k$, the assumed nonzero
communication and sensing projections imply
\begin{equation}
p_k(t\mathbf{v}_k)\rightarrow1,
\qquad
\mathrm{CRB}_{\theta}(t\mathbf{v}_k)\rightarrow0
\end{equation}
as $t\rightarrow\infty$. Consequently, the left-hand side of
\eqref{eq:joint_lyapunov} converges to
$L_{\mathrm{cl},k}
+\operatorname{Tr}(\mathbf{P}\boldsymbol{\Sigma}_p)$.
Since
\eqref{eq:controller_feasibility} provides a strict positive
margin, continuity guarantees the existence of a sufficiently
large but finite $t$ for which
$\mathbf{w}_k=t\mathbf{v}_k$ satisfies
\eqref{eq:joint_lyapunov}.

We first show that the expected Lyapunov function value is bounded at each time step. Under the proposed joint ISCC optimization framework, the transmit beamforming vector $\mathbf{w}_k$ is designed to satisfy the state-dependent Lyapunov constraint at each time slot. Thus, the conditional expectation of the Lyapunov function value at the next time slot satisfies
\begin{equation}
\mathbb{E}[V(\mathbf{q}_{k+1})\mid\mathbf{q}_k] \le \rho(\|\mathbf{q}_k\|)V(\mathbf{q}_k) + C_0.
\end{equation}

Since $\tanh(\|\mathbf{q}_k\|) \ge 0$, the contraction factor satisfies $\rho(\|\mathbf{q}_k\|) \le \rho_0 < 1$ for all $\mathbf{q}_k$. Substituting this bound into the conditional inequality yields
\begin{equation}
\mathbb{E}[V(\mathbf{q}_{k+1})\mid\mathbf{q}_k] \le \rho_0 V(\mathbf{q}_k) + C_0.
\end{equation}

Taking expectations on both sides and applying the law of total expectation yield the following recursive inequality:
\begin{equation}
\mathbb{E}[V(\mathbf{q}_{k+1})] \le \rho_0 \mathbb{E}[V(\mathbf{q}_k)] + C_0.
\end{equation}
Iterating the above linear recursion from the initial state $\mathbf{q}_0$ to time step $k$, it can be readily verified that
\begin{equation}
\mathbb{E}[V(\mathbf{q}_k)] \le \rho_0^k \mathbb{E}[V(\mathbf{q}_0)] + C_0 \sum_{i=0}^{k-1} \rho_0^i.
\end{equation}

We next evaluate the steady-state behavior of the closed-loop system. Taking the limit superior as $k \to \infty$, the transient term $\rho_0^k \mathbb{E}[V(\mathbf{q}_0)]$ asymptotically vanishes since $\rho_0 \in (0,1)$. Using the sum of the infinite geometric series, the steady-state energy bound is given by
\begin{equation}
\limsup_{k \to \infty} \mathbb{E}[V(\mathbf{q}_k)] \le \frac{C_0}{1-\rho_0}.
\end{equation}

Finally, taking into account the Rayleigh quotient property of the positive definite matrix $\mathbf{P}$, we have $\lambda_{\min}(\mathbf{P})\|\mathbf{q}_k\|^2 \le \mathbf{q}_k^T \mathbf{P} \mathbf{q}_k = V(\mathbf{q}_k)$. Taking the unconditional expectation and substituting it into the steady-state energy bound, we obtain
\begin{equation}
\limsup_{k \to \infty} \mathbb{E}[\|\mathbf{q}_k\|^2] \le \frac{1}{\lambda_{\min}(\mathbf{P})} \frac{C_0}{1-\rho_0},
\end{equation}
which implies that the system state sequence is mean-square ultimately bounded. This completes the proof. \hfill $\blacksquare$



\ifCLASSOPTIONcaptionsoff
  \newpage
\fi



\bibliographystyle{IEEEtran}
\bibliography{bib/IEEEabrv,bib/references}
%

%





\end{document}